\documentclass[paper=A4,11pt,DIV=12,headings=big]{scrartcl}
\pdfoutput=1
\usepackage{graphicx}
\usepackage{tabularx}
\usepackage{array}
\usepackage{amsmath}
\usepackage{amssymb}
\usepackage{color}
\usepackage{longtable}
\usepackage{verbatim}
\usepackage{cite}
\usepackage{bbm}
\usepackage{multirow}
\usepackage[normalem]{ulem}
\usepackage[bookmarks=false]{hyperref}

\addtokomafont{disposition}{\rmfamily\boldmath}

\newcommand{\beq}{\begin{equation}}
\newcommand{\eeq}{\end{equation}}
\newcommand{\ba}{\begin{array}}
\newcommand{\ea}{\end{array}} 
\newcommand{\beqa}{\begin{eqnarray}}
\newcommand{\eeqa}{\end{eqnarray}}

\def \eff{\rm eff}

\newcommand{\bsmm}{B_s\to\mu^+\mu^-}

\begin{document}

\begin{flushright}
\begin{tabular}{l}
FERMILAB-PUB-12-257-T
\end{tabular}
\end{flushright}
\vskip1.5cm

\begin{center}
{\LARGE \bf \boldmath Cornering New Physics in $b \to s$ Transitions}\\[0.8 cm]
{\large
Wolfgang~Altmannshofer$^{a}$ and David M. Straub$^{b}$} \\[0.5 cm]
\small
$^a$ {\em Fermi National Accelerator Laboratory, P.O. Box 500, Batavia, IL 60510, USA}\\[0.1cm]
$^b${\em Scuola Normale Superiore and INFN, Piazza dei Cavalieri 7, 56126 Pisa, Italy}
\end{center}

\bigskip
\begin{abstract}\noindent
We derive constraints on Wilson coefficients of dimension-six effective operators probing the  $b\to s$ transition, using recent improved measurements of the rare decays $\bsmm$, $B\to K\mu^+\mu^-$ and $B\to K^*\mu^+\mu^-$ and including all relevant observables in inclusive and exclusive decays. We consider operators present in the SM as well as their chirality-flipped counterparts and scalar operators.
We find good agreement with the SM expectations. Compared to the situation before winter 2012, we find significantly more stringent constraints on the chirality-flipped coefficients due to complementary constraints from $B\to K\mu^+\mu^-$ and $B\to K^*\mu^+\mu^-$ and due to the LHCb measurement of the angular observable $S_3$ in the latter decay.
We also list the full set of observables sensitive to new physics in the low recoil region of $B\to K^*\mu^+\mu^-$.
\end{abstract}

\section{Introduction}

Rare $B$ meson decays probing the flavour-changing neutral current $b\to s$ transition are sensitive to physics beyond the Standard Model (SM). Independently of a concrete model, 
the effect of new degrees of freedom that are sufficiently heavy compared to the decaying $B$ mesons
can be described by the modification of Wilson coefficients of local operators in an effective Hamiltonian of the form
\begin{equation}
\label{eq:Heff}
{\cal H}_{\eff} = - \frac{4\,G_F}{\sqrt{2}} V_{tb}V_{ts}^* \frac{e^2}{16\pi^2}
\sum_i
(C_i O_i + C'_i O'_i) + \text{h.c.}~.
\end{equation}
Since the set of relevant operators and the dependence on the Wilson coefficients can be different for the various experimental observables probing this Hamiltonian, a combined analysis of all available experimental constraints is mandatory to obtain meaningful bounds on the individual coefficients and to determine the room left for new physics (NP). Since the startup of the LHC, such constraints have become significantly more stringent, thanks to improved measurements of exclusive decays. In the last few months, the LHCb and BaBar collaboration have presented updated analyses of branching ratios and angular observables in $B\to K^{(*)}\mu^+\mu^-$, while the LHCb, ATLAS and CMS experiment have presented strong bounds on the branching ratio of the very rare $B_s\to\mu^+\mu^-$ decay. The purpose of this paper is to update the model-independent constraints on Wilson coefficients first presented in \cite{Altmannshofer:2011gn}.
As in the previous analysis, we go beyond comparable recent studies 
\cite{Bobeth:2010wg,DescotesGenon:2011yn,Bobeth:2011gi,Bobeth:2011nj,Beaujean:2012uj,Becirevic:2012fy} by considering the SM operator basis as well as their chirality-flipped counterparts and by including the most complete set of observables in inclusive and exclusive decays.
With respect to \cite{Altmannshofer:2011gn}, we emphasize the most important improvements:
\begin{itemize}
\item We take into account the new data on $B\to K^{(*)}\mu^+\mu^-$ and $B_s\to\mu^+\mu^-$ released by LHCb and BaBar in 2012.
\item We include the decays $B\to K\mu^+\mu^-$ and $B_s\to\mu^+\mu^-$ in the determination of Wilson coefficients constraints.
\item We list the full set of $B\to K^*\mu^+\mu^-$ angular observables sensitive to NP at high dilepton invariant mass $q^2$, pointing out that there are 5 observables only sensitive to right-handed currents. This is particularly relevant for $S_3$ recently measured by LHCb and CDF, which we now include in our analysis also at high $q^2$.
\item We discuss the impact of the direct CP asymmetry in $B\to X_s\gamma$ on the constraints, considering the sizable theory uncertainties due to long-distance contributions.
\end{itemize}
Next, we will discuss the relevant observables and their dependence on NP in sec.~\ref{sec:obs}, focusing on the improvements relative to \cite{Altmannshofer:2011gn}. The resulting constraints on individual Wilson coefficients are presented in sec.~\ref{sec:modelind}. Sec.~\ref{sec:prospects} is devoted to a discussion of the potential size of NP effects in observables yet to be measured, which is useful to assess the prospects of future measurements.
We summarize our results in sec.~\ref{sec:concl}.
Finally, appendix~\ref{app:exp} contains a detailed discussion of our method to average measurements from different experiments and compares these averages to our theoretical predictions.

\section{Observables}\label{sec:obs}

We consider NP contributions to the operators
\begin{align}
\label{eq:O7}
O_7^{(\prime)} &= \frac{m_b}{e}
(\bar{s} \sigma_{\mu \nu} P_{R(L)} b) F^{\mu \nu},
&
O_8^{(\prime)} &= \frac{g_s m_b}{e^2}
(\bar{s} \sigma_{\mu \nu} T^a P_{R(L)} b) G^{\mu \nu \, a},
\nonumber\\
O_9^{(\prime)} &= 
(\bar{s} \gamma_{\mu} P_{L(R)} b)(\bar{\ell} \gamma^\mu \ell)\,,
&
O_{10}^{(\prime)} &=
(\bar{s} \gamma_{\mu} P_{L(R)} b)( \bar{\ell} \gamma^\mu \gamma_5 \ell)\,,
\nonumber\\
O_S^{(\prime)} &= 
\frac{m_b}{m_{B_s}} (\bar{s} P_{R(L)} b)(  \bar{\ell} \ell)\,,
&
O_P^{(\prime)} &=
\frac{m_b}{m_{B_s}} (\bar{s} P_{R(L)} b)(  \bar{\ell} \gamma_5 \ell)\,,
\end{align}
and define all the Wilson coefficients at a matching scale of 160~GeV. In the SM, the primed coefficients as well as $C_{S,P}$ are negligibly small. For numerical values of the Wilson coefficients we refer to the appendix of \cite{Altmannshofer:2011gn}. Our normalization choice for the scalar and pseudoscalar operators is chosen to make their Wilson coefficients dimensionless and scale invariant. We note that, by factoring out the CKM elements in the definition of (\ref{eq:Heff}), a non-zero imaginary part for a Wilson coefficient corresponds to CP violation beyond the CKM phase.

We consider all relevant observables in inclusive and exclusive $B$ meson decays that are sensitive to the Wilson coefficients of the operators in~(\ref{eq:O7}).
In particular, in addition to the observables considered in~\cite{Altmannshofer:2011gn} -- the branching ratio of the inclusive $B \to X_s \gamma$ decay, the time dependent CP asymmetry in the exclusive $B \to K^* \gamma$ decay, the branching ratio of the inclusive $B \to X_s \ell^+\ell^-$ decay, the branching ratio and angular observables in the exclusive $B \to K^* \mu^+\mu^-$ decay and the branching ratio of the rare $B_s \to \mu^+\mu^-$ decay -- we also include the branching ratio of the exclusive $B \to K \mu^+\mu^-$ decay, that provides valuable complementary information on the right-handed Wilson coefficients, and the direct CP asymmetry in $B \to X_s \gamma$, $A_{\rm CP}(b \to s \gamma)$.

Our treatment of BR($B \to X_s \gamma$) and $B \to K^* \gamma$ exactly follows our analysis in~\cite{Altmannshofer:2011gn}. We slightly refine our treatment of $B \to X_s \ell^+ \ell^-$ by adding a 10\% theory uncertainty due to the cut on the invariant mass of the hadronic final state~\cite{Lee:2005pwa}. In view of the current large experimental uncertainty, the impact on our conclusions is marginal.
Improvements in our treatment of $B \to K^* \mu^+\mu^-$ and $B_s \to \mu^+\mu^-$ on both the theory and experimental side, as well as our analyses of $B \to K \mu^+\mu^-$ and $A_{\rm CP}(b \to s \gamma)$, are described in detail below.

\begin{table}[tp]
\renewcommand{\arraystretch}{1.2}
\begin{center}
\begin{tabular}{lcclcl}
\hline   
Observable & $q^2$ & Experiment &  &  SM prediction &  \\  
\hline
$10^4 \times$BR$(B \to X_s \gamma)$ && $3.55\pm 0.26$ & \cite{Asner:2010qj} & $3.15\pm 0.23$ & \cite{Misiak:2006zs} \\
$A_\text{CP}(b \to s \gamma)$ && $(-1.2 \pm 2.8)\%$ & \cite{Asner:2010qj} & $( 1.0\pm1.6 )\%$ & \cite{Benzke:2010tq} \\
$S_{K^*\gamma}$ && $-0.16 \pm 0.22$ & \cite{Asner:2010qj} & $(-2.3\pm1.6)\%$ & \cite{Ball:2006eu} \\
$10^9\times\text{BR}(\bsmm)$ && $2.4 \pm 1.6$ & \cite{Aaij:2012ac,Chatrchyan:2012rg,Aad:2012pn,Aaltonen:2011fi} & $3.32\pm0.17$ & \\
\hline
\multirow{2}{*}{$10^6 \times$BR$(B \to X_s \ell^+ \ell^-)$}&${[1,6]}$ & $1.63 \pm  0.50$ & \multirow{2}{*}{\cite{Aubert:2004it,Iwasaki:2005sy}} & $1.59 \pm 0.19$ & \cite{Huber:2005ig,Lee:2005pwa} \\
&${>14.4}$ & $0.43 \pm 0.12$ &  & $0.23 \pm 0.07$ & \cite{Hurth:2008jc} \\
\hline
\multirow{3}{*}{$10^7 \times$BR$(B\to K\ell^+ \ell^-)$}&${[1,6]}$ & $1.28 \pm 0.12$ & \multirow{3}{*}{\cite{Aaltonen:2011qs,:2009zv,Collaboration:2012vw,LHCbisospin}} & $1.29 \pm 0.30$ & \\
&${[14.18,16]}$ & $0.41 \pm 0.06$ &  & $0.43 \pm 0.10$ & \\
&${[16,22.9]}$ & $0.49 \pm 0.07$ & & $0.86 \pm 0.20$ & \\
\hline
\multirow{3}{*}{$10^7 \times$BR$(B \to K^* \ell^+ \ell^-)$}&${[1,6]}$ & $1.90 \pm 0.19$ & \multirow{3}{*}{\cite{Aaltonen:2011qs,:2009zv,Collaboration:2012vw,LHCb-CONF-2012-008,LHCbisospin}} & $2.49\pm0.64$ & \\
&${[14.18,16]}$ & $1.12 \pm 0.11$ &  & $1.13 \pm 0.41$ & \\
&${[16,19]}$ & $1.27 \pm 0.12$ &  & $1.34 \pm 0.62$ & \\
\hline
\multirow{3}{*}{$\langle F_L\rangle(B \to K^* \ell^+ \ell^-)$}&${[1,6]}$ & $0.63\pm0.06$ & \multirow{3}{*}{\cite{Aaltonen:2011ja,:2009zv,akar,LHCb-CONF-2012-008}} & $0.77\pm0.04$ & \\
&${[14.18,16]}$ & $0.34\pm0.06$ &  & $0.37\pm0.20$ & \\
&${[16,19]}$ & $0.32\pm0.06$ &  & $0.34\pm0.25$ & \\
\hline
\multirow{3}{*}{$\langle A_\text{FB} \rangle(B \to K^* \ell^+ \ell^-)$}&${[1,6]}$ & $0.09 \pm 0.05$ & \multirow{3}{*}{\cite{Aaltonen:2011ja,:2009zv,akar,LHCb-CONF-2012-008}} & $0.03\pm0.02$ & \\
&${[14.18,16]}$ & $-0.46\pm0.05$ &  & $-0.41\pm0.13$ & \\
&${[16,19]}$ & $-0.39\pm0.06$ &  & $-0.35\pm0.13$ & \\
\hline
\multirow{3}{*}{$\langle S_3 \rangle(B \to K^* \ell^+ \ell^-)$}&${[1,6]}$ & $0.05 \pm 0.08$ & \multirow{3}{*}{\cite{Aaltonen:2011ja,LHCb-CONF-2012-008}} & $(-0.3\pm1.1)\, 10^{-2}$ & \\
&${[14.18,16]}$ & $0.02\pm0.08$ &  & $-0.14\pm0.08$ & \\
&${[16,19]}$ & $-0.21\pm0.08$ &  & $-0.22\pm0.10$ & \\
\hline
\multirow{3}{*}{$\langle A_9 \rangle(B \to K^* \ell^+ \ell^-)$}&${[1,6]}$ & $0.09\pm0.39$
 & \multirow{3}{*}{\cite{Aaltonen:2011ja}} & $(1.5 \pm 2.4)\, 10^{-4}$ & \\
&${[14.18,16]}$ & $0.18\pm0.25$
 &  & $(-0.7 \pm 1.7)\, 10^{-5}$ & \\
&${[16,19]}$ & $-0.20\pm0.34$
 &  & $(-0.5 \pm 1.2)\, 10^{-5}$ & \\
\hline
\end{tabular}
\end{center}
\caption{Experimental averages and SM predictions for the observables used in the fit. SM predictions that lack a reference are based on our calculations.}
\label{tab:obs} 
\end{table}

\subsection{$B_s \to \mu^+\mu^-$}\label{sec:bsmm}

Recently, it has been pointed out~\cite{deBruyn:2012wj,deBruyn:2012wk} that the large width difference in the $B_s$ system, $y_s = \tau_{B_s} \Delta \Gamma_s / 2 = (8.8 \pm 1.4)\%$~\cite{LHCb-CONF-2012-002},
has to be taken into account when comparing theory predictions of BR$(B_s \to \mu^+\mu^-)$ with the experimental results. The corresponding correction is not universal but depends on possible NP contributions to $B_s \to \mu^+\mu^-$. In presence of NP one has
\begin{equation} \label{eq:BRbsmm}
\frac{{\rm BR}(B_s \to \mu^+\mu^-)}{{\rm BR}(B_s \to \mu^+\mu^-)_{\rm SM}} \simeq
\Big( |S|^2 + |P|^2 \Big) \times \left( 1 + y_s \frac{{\rm Re}(P^2) - {\rm Re}(S^2) }{|S|^2 + |P|^2} \right) \Big/ (1 + y_s) ~,
\end{equation}
where
\begin{equation} \label{eq:SP}
S = \frac{m_{B_s}}{2 m_\mu} \frac{(C_S - C_S')}{C^{\rm SM}_{10}} \sqrt{\left( 1 - \frac{4 m_\mu^2}{m_{B_s}^2} \right)},
\qquad
P = \frac{m_{B_s}}{2 m_\mu} \frac{(C_P - C_P')}{C^{\rm SM}_{10}} +
\frac{(C_{10} - C_{10}')}{C^{\rm SM}_{10}} ~.
\end{equation}
In (\ref{eq:BRbsmm}), BR$(B_s \to \mu^+\mu^-)_{\rm SM}$ and BR$(B_s \to \mu^+\mu^-)$ refer to the SM and NP predictions for the branching ratio extracted from an untagged rate that can directly be compared to experiment~\cite{deBruyn:2012wj,deBruyn:2012wk}. Using an average of the most recent lattice determinations of the $B_s$ meson decay constant~\cite{McNeile:2011ng,Bazavov:2011aa,Na:2012kp}, $f_{B_s} = (227 \pm 4)$MeV~\cite{Davies:2012qf}, we find\footnote{Neglecting the $y_s$ correction we find BR$(B_s \to \mu^+\mu^-)_{\rm SM} = (3.05 \pm 0.15) \times 10^{-9}$, which corresponds to the CP averaged branching ratio.}
\begin{equation}
{\rm BR}(B_s \to \mu^+\mu^-)_{\rm SM} = (3.32 \pm 0.17) \times 10^{-9}~.
\end{equation}

The experimental sensitivities start to close in on the SM value. While CDF~\cite{Aaltonen:2011fi} finds a small excess in $B_s \to \mu^+ \mu^-$ candidates and gives a two-sided limit on the BR$(B_s \to \mu^+\mu^-)$ at 95\% C.L., LHCb~\cite{Aaij:2012ac}, CMS~\cite{Chatrchyan:2012rg} and ATLAS~\cite{Aad:2012pn} quote at 95\% C.L. only upper bounds on the branching ratio. At the 2$\sigma$ level all results are consistent and we perform a naive combination using the available $\Delta \chi^2$ distribution for BR$(B_s \to \mu^+ \mu^-)$ from CDF as well as the CL$_s$ distributions from ATLAS, CMS and LHCb. We obtain
\begin{equation}
{\rm BR}(B_s \to \mu^+ \mu^-)_{\rm exp} = (2.4 \pm 1.6) \times 10^{-9} ~, 
\label{eq:bsmmexp}
\end{equation}
which is in good agreement with the SM expectation.
While this clearly leads to strong constraints on models with potentially large contributions to the scalar or pseudoscalar operators, contributions to the Wilson coefficients $C_{10},C_{10}'$ were known even before the measurement (\ref{eq:bsmmexp}) to be unable to enhance the branching ratio above about $5.6\times10^{-9}$ due to constraints from other rare semi-leptonic decays, as shown in \cite{Altmannshofer:2011gn}. As a consequence, the experimental precision for this decay is only now starting to become sensitive to models modifying the branching ratio by means of the semi-leptonic operators $O_{10}$ and $O_{10}^\prime$.

\subsection{$B \to K^* \mu^+\mu^-$}\label{sec:BKstar}

The $B \to K^* \mu^+\mu^-$ decay offers a multitude of observables that are sensitive to NP effects (see e.g.~\cite{Bobeth:2008ij,Altmannshofer:2008dz,Bobeth:2010wg,Matias:2012xw}). 
Updated measurements of the branching ratio and angular observables in $B \to K^* \mu^+\mu^-$ have recently been presented by BaBar \cite{Collaboration:2012vw,akar} and LHCb \cite{LHCb-CONF-2012-008,LHCbisospin}.
In addition to the branching ratio, forward-backward asymmetry $A_\text{FB}$ and the $K^*$ longitudinal polarization fraction $F_L$, LHCb also strongly improved the constraints on the CP-averaged angular observable $S_3$
first constrained by CDF \cite{Aaltonen:2011ja}, which is tiny in the SM but highly sensitive to NP in right-handed currents.
Compared to \cite{Altmannshofer:2011gn}, we now also include the $S_3$
constraint at high $q^2$ and find it to give a relevant constraint on the primed Wilson coefficients.
We also include the angular CP asymmetry $A_9$, that is very sensitive to CP violation in right-handed currents, both at low and high $q^2$. Currently, only weak bounds on this observable exist from CDF~\cite{Aaltonen:2011ja} and their impact on our results is small.
The impact of a future measurement of $A_9$ by LHCb will be discussed at the end of section \ref{sec:prospects}.

\begin{table}
\begin{center}
\renewcommand{\arraystretch}{1.2}
\begin{tabular}{lll}
\hline
Obs. & low $q^2$ & high $q^2$ \\
\hline
BR & $C_{7,9,10},C_{7,9,10}'$ & $C_{9,10},C_{9,10}'$ \\
$F_L$ & $C_{7,9}, C_{7,9,10}'$ & $C_{9,10}'$\\
$S_3$ & $C_{7,10}'$ & $C_{9,10}'$\\
$S_4$ & $C_{7,10},C_{7,10}'$ & $C_{9,10}'$\\
$S_5$ & $C_{7,9},C_{7,10}'$& $C_9, C_{9,10}'$ \\
$A_\text{FB}$ & $C_7,C_9$ & $C_{9,10}, C_{9,10}'$ \\
$A_7$ & $C_{7,10},C_{7,10}'$ & -- \\
$A_8$ & $C_{7,9},C_{7,9,10}'$ & $C_{9,10}'$\\
$A_9$ & $C_{7,10}'$ & $C_{9,10}'$\\
\hline
\end{tabular}
\end{center}
\caption{Observables in the angular distribution of $B\to K^*\mu^+\mu^-$ and Wilson coefficients they are most sensitive to in the low and high $q^2$ regions.}
\label{tab:sens}
\end{table}

Concerning the high $q^2$ region, we also emphasize that with our operator basis, eq.~(\ref{eq:O7}), there are more independent observables at high $q^2$ than in the case of the SM operator basis. For example, while it has been shown in \cite{Bobeth:2010wg} that $F_L$ is ``short-distance free'' at high $q^2$ in the case of the SM operator basis and can be used to extract form factor ratios from the data \cite{Hambrock:2012dg}, this is no longer true in the presence of NP contributions to $C_{7,9,10}'$. Consequently, $F_L$ at high $q^2$ can be considered a probe of right-handed currents. The same is true for the observable $S_4$ defined in \cite{Altmannshofer:2008dz} and for the CP asymmetry $A_8$. The CP asymmetry $A_7$ is instead ``short-distance free'' at high $q^2$ even in the presence of chirality-flipped coefficients, while $S_5$ at high $q^2$ is sensitive to NP even with SM-operators only. The sensitivity to NP in the various observables is summarized for convenience in table~\ref{tab:sens}.

On the experimental side, both
LHCb and BaBar observe some hints for a non-zero isospin asymmetry.
However, since the theoretical expectation for the isospin asymmetry is rather limited compared to the experimental uncertainties even in the presence of NP \cite{Feldmann:2002iw}, we feel justified in neglecting it
in the following and use both the $B^+$ and $B^0$ data in our averages. Details on the averaging procedure and the comparison of experiment vs.\ theory are discussed in appendix~\ref{app:exp}.

Finally, we note two minor numerical refinements of the SM predictions compared to \cite{Altmannshofer:2011gn}:
\begin{itemize}
\item We use an updated value for the neutral $B$ meson decay constant $f_B=190(4)$~MeV \cite{Davies:2012qf} relevant for the normalization of $B\to K^*$ form factors, the effect being an increased SM prediction for the branching ratio in the low $q^2$ bin (but small with respect to the theory uncertainty);
\item For the extrapolation of form factors to high $q^2$, where we use the results of \cite{Bharucha:2010im}, we now take into account the error correlations between the fitted expansion parameters. This leads to a very small increase in the estimated theory uncertainties.
\end{itemize}
Our predictions for all the relevant observables at low and high $q^2$ are listed in table~\ref{tab:obs}.
As we take into account error correlations for the $B\to K^*$ form factors in the low $q^2$ region~\cite{Altmannshofer:2008dz}, some of our theory predictions are considerably more precise than the values in~\cite{Bobeth:2010wg}.

\subsection{$B \to K \mu^+\mu^-$}\label{sec:BK}

The branching ratio of the $B \to K \mu^+\mu^-$ decay has been measured by the BaBar, Belle, CDF and LHCb collaborations \cite{Aaltonen:2011qs,:2009zv,Collaboration:2012vw,LHCbisospin}. Its impact on constraints on the Wilson coefficients $C_{7,9,10}$ has recently been analyzed in \cite{Bobeth:2011nj,Beaujean:2012uj}. Here, we include the BR$(B\to K\mu^+ \mu^-)$ in our analysis, discussing also the chirality-flipped Wilson coefficients $C_{7,9,10}'$.
Since the $B\to K$ transition does not receive contributions from an axial vector current, the primed Wilson coefficients enter the $B\to K\mu^+\mu^-$ observables always in conjunction with their unprimed counterparts as $(C_i+C_i')$.
This is complementary to the $B\to K^*\mu^+\mu^-$ decay and is useful to constrain the chirality-flipped operators. Neglecting lepton mass effects and (pseudo)scalar operators, the differential branching ratio can be written as \cite{Bobeth:2007dw}
\begin{equation}
\frac{d\text{BR}}{dq^2}=
\frac{\tau_BG_F^2\alpha_e^2|V_{tb}V_{ts}^*|^2}{2^9\pi^5m_B^3}
\frac{\lambda^{3/2}}{3}
\left(|F_A|^2 +|F_V|^2\right),
\end{equation}
where
\begin{equation} \label{eq:FAFV}
F_A=(C_{10}+C_{10}')f_+(q^2)\,,
\qquad
F_V=(C_{9}^\text{eff}+C_9')f_+(q^2)+\frac{2m_b}{m_B+m_K}(C_7^\text{eff}+C_7')f_T(q^2)\,,
\end{equation}
\begin{equation}
\lambda = m_B^4 + m_K^4 + q^4 - 2(m_B^2 m_K^2 + m_B^2 q^2 + m_K^2 q^2)\,.
\end{equation}
In~(\ref{eq:FAFV}) the Wilson coefficients are evaluated at the scale of the $b$ quark, $\mu \simeq m_b$.
The effective Wilson coefficients $C_{7,9}^\text{eff}$ are defined for instance in \cite{Altmannshofer:2008dz}. In addition, we take into account non-factorizable $O(\alpha_s)$ corrections proportional to form factors \cite{Beneke:2001at}.
For the $B\to K$ form factors, we use the results of \cite{Bharucha:2010im}, which are based on a fit to a LCSR calculation valid at low $q^2$ and lattice results valid at high $q^2$. We take into account all the error correlations among the fit parameters given in an unpublished updated version of \cite{Bharucha:2010im}\footnote{We thank Aoife Bharucha for providing us with the correct covariance matrices.}.
In view of the sizable resulting form factor uncertainties, we neglect the impact of possible higher-order non-perturbative corrections, which are estimated to be at the few percent level \cite{Bartsch:2009qp,Beylich:2011aq}.
Similarly to the case of $B\to K^*\mu^+\mu^-$, we ignore isospin breaking effects, which are expected to be well below the experimental sensitivity, and consequently use the average of $B^0$ and $B^+$ data for the experimental branching ratios.
Our resulting SM predictions for the branching ratio in the different $q^2$ bins are shown in table~\ref{tab:obs}.

\subsection{$A_{\rm CP}(b \to s \gamma)$}\label{sec:ACP}

The direct CP asymmetry in the $B \to X_s \gamma$ decay probes CP phases in the Wilson coefficients $C_7$ and $C_7^\prime$ of the magnetic operators. It arises first at NLO and therefore depends also in a non-trivial way on the Wilson coefficients $C_8$ and $C_8^\prime$ of the chromo-magnetic operators. This is in contrast to all the other observables sensitive to $C_7^{(\prime)}$ that we consider in this work that all depend on the same combination of $C_7^{(\prime)}$ and $C_8^{(\prime)}$.

The experimental world average for $A_{\rm CP}(b \to s \gamma)$ is dominated by data from Belle~\cite{Nishida:2003yw} and BaBar~\cite{Aubert:2008be} and reads~\cite{Asner:2010qj}
\begin{equation}
 A_{\rm CP}(b \to s \gamma)_\text{exp} = (-1.2 \pm 2.8) \% ~,
\end{equation}
which is consistent with CP conservation.

The SM prediction for $A_{\rm CP}(b \to s \gamma)$ suffers from large hadronic uncertainties~\cite{Benzke:2010tq}
\begin{equation}
 -0.6\% < A_{\rm CP}(b \to s \gamma)_\text{exp} < 2.8\% ~,
\end{equation}
which makes it difficult to identify small NP contributions in this observable. Nonetheless, $A_{\rm CP}(b \to s \gamma)$ can be used to constrain large CP violating NP contributions
which would otherwise be allowed by measurements of CP-averaged quantities like branching ratios.

In our numerical analysis, we use the expressions given in~\cite{Benzke:2010tq} that we extend to include also the right-handed Wilson coefficients $C_7^\prime$ and $C_8^\prime$. To estimate the theoretical uncertainty in presence of NP, we consider the uncertainties in the so-called resolved photon contributions given in~\cite{Benzke:2010tq},
treating the model estimates for the $\tilde \Lambda$ parameters as Gaussian $1\sigma$ ranges, as well as
the uncertainties in the remaining contributions coming from scale variation and $m_c/m_b$ given in~\cite{Hurth:2003dk}. We add all uncertainties in quadrature. Using this procedure, we obtain the SM prediction quoted in table~\ref{tab:obs}.

\section{Model-independent constraints on Wilson coefficients}\label{sec:modelind}

Combining all the experimental averages and theory predictions of the observables listed in table~\ref{tab:obs}, we construct a $\chi^2$ function and use it to obtain model-independent constraints on the Wilson coefficients. Our statistical treatment follows \cite{Altmannshofer:2011gn}, with one minor refinement: while previously treating all theory uncertainties as uncorrelated, we now treat the uncertainties of the two adjacent high-$q^2$ bins for all observables in $B\to K^*\mu^+\mu^-$ and $B\to K\mu^+\mu^-$ as 100\% correlated, which we found to be approximately fulfilled for generic values of Wilson coefficients and which we checked to lead to slightly more conservative (looser) constraints than assuming them to be uncorrelated.

\begin{figure}[p]
\centering
\includegraphics[width=\textwidth]{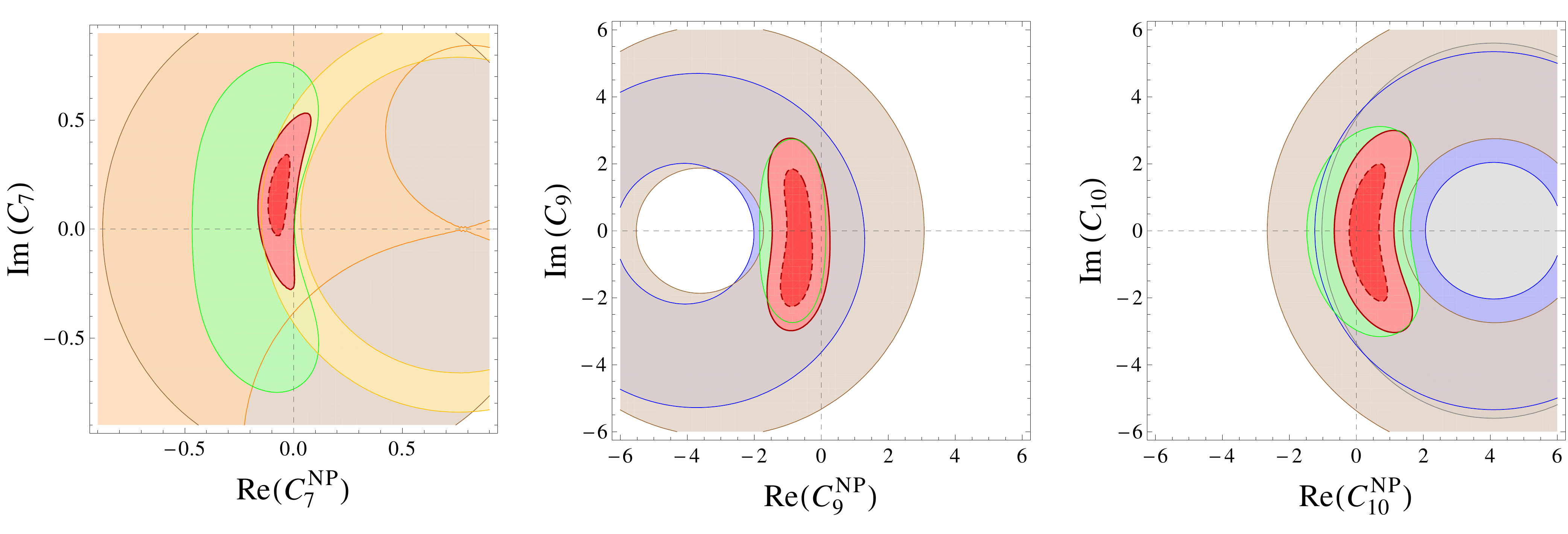}
\caption{Individual $2\sigma$ constraints on the unprimed Wilson coefficients from
$B\to X_s\ell^+\ell^-$ (brown), BR($B\to X_s\gamma$) (yellow), $A_\text{CP}(b\to s\gamma)$ (orange), $B\to K^*\gamma$ (purple), $B\to K^*\mu^+\mu^-$ (green), $B\to K\mu^+\mu^-$ (blue) and $B_s\to\mu^+\mu^-$ (gray) as well as combined 1 and $2\sigma$ constraints (red).
}
\label{fig:bandplots1}
\end{figure}

\begin{figure}[p]
\centering
\includegraphics[width=\textwidth]{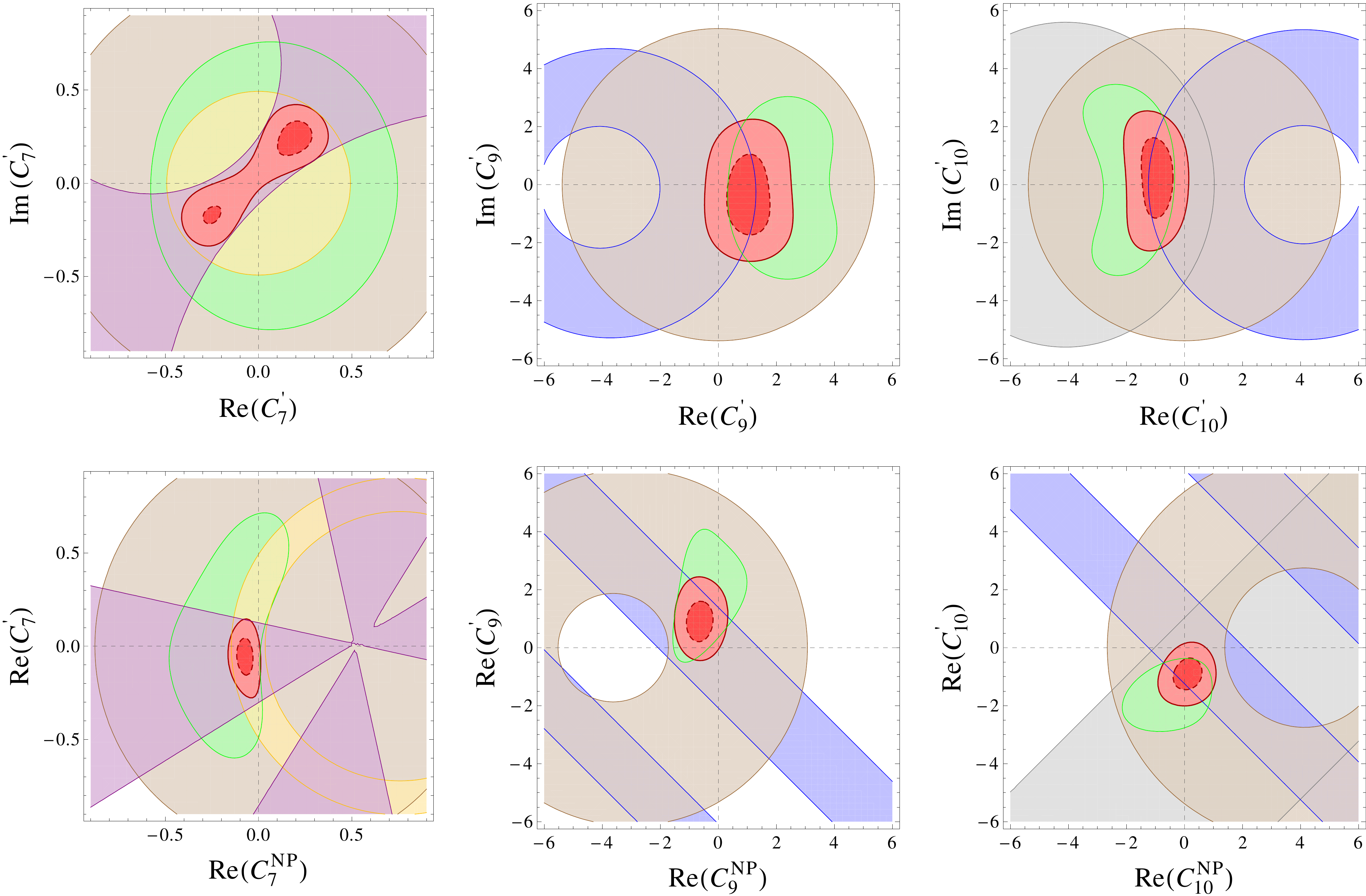}
\caption{Individual $2\sigma$ constraints on the primed Wilson coefficients as well as combined 1 and $2\sigma$ constraints. Same colour coding as in figure~\ref{fig:bandplots1}.
}
\label{fig:bandplots2}
\end{figure}

We first discuss the constraints on single complex coefficients or on pairs of real Wilson coefficients. Figures~\ref{fig:bandplots1} and \ref{fig:bandplots2} shows contours of $\Delta \chi^2=4$ ($\sim2\sigma$ constraints) from individual processes and $\Delta \chi^2=1,4$ ($\sim1,2\sigma$ constraints) from the combined constraints.
We make the following observations.
\begin{itemize}
\item At the 95\% C.L., all Wilson coefficients are compatible with their SM values. For the total $\chi^2$ for the SM values of the Wilson coefficients we find $\chi^2/N_{\rm dof} = 21.8/24$. This value improves only slightly in presence of NP.
\item For the coefficients present in the SM, i.e. $C_7$, $C_9$ and $C_{10}$, as well as for $C_9^\prime$ and $C_{10}^\prime$, the constraints on the imaginary part are looser than on the real part.
\item For the Wilson coefficients $C_{10}^{(\prime)}$, the new constraint on $B_s\to\mu^+\mu^-$ is starting to become competitive with the constraints from $B\to K^{(*)}\mu^+\mu^-$.
\item The constraints on $C_9'$ and $C_{10}'$ from $B\to K\mu^+\mu^-$ and $B\to K^*\mu^+\mu^-$ are complementary and lead to a more constrained region, and better agreement with the SM, than with $B\to K^*\mu^+\mu^-$ alone.
\item A second allowed region in the $C_7$-$C_7'$ plane characterized by large positive contributions to both coefficients, which was found to be allowed e.g. in \cite{Altmannshofer:2011gn,DescotesGenon:2011yn}, is now disfavoured at 95\% C.L. by the new $B\to K^*\mu^+\mu^-$ data, in particular the forward-backward asymmetry.
\item The $A_\text{CP}(b \to s \gamma)$ constraint limits the size of allowed imaginary contributions to $C_7$. Its impact on $C_7'$ is marginal, so we refrain from showing it.
\end{itemize}
The second point above can be understood from the fact that in the branching ratios and CP averaged angular observables giving the strongest constraints, only NP contributions aligned in phase with the SM can interfere with the SM contributions. As a consequence, NP with non-standard CP violation is in fact constrained more weakly than NP where CP violation stems only from the CKM phase. This highlights the need for improved measurements of CP asymmetries directly sensitive to non-standard phases, such as the T-odd CP asymmetries in $B\to K^*\mu^+\mu^-$.

\begin{figure}[tbp]
\centering
\includegraphics[width=\textwidth]{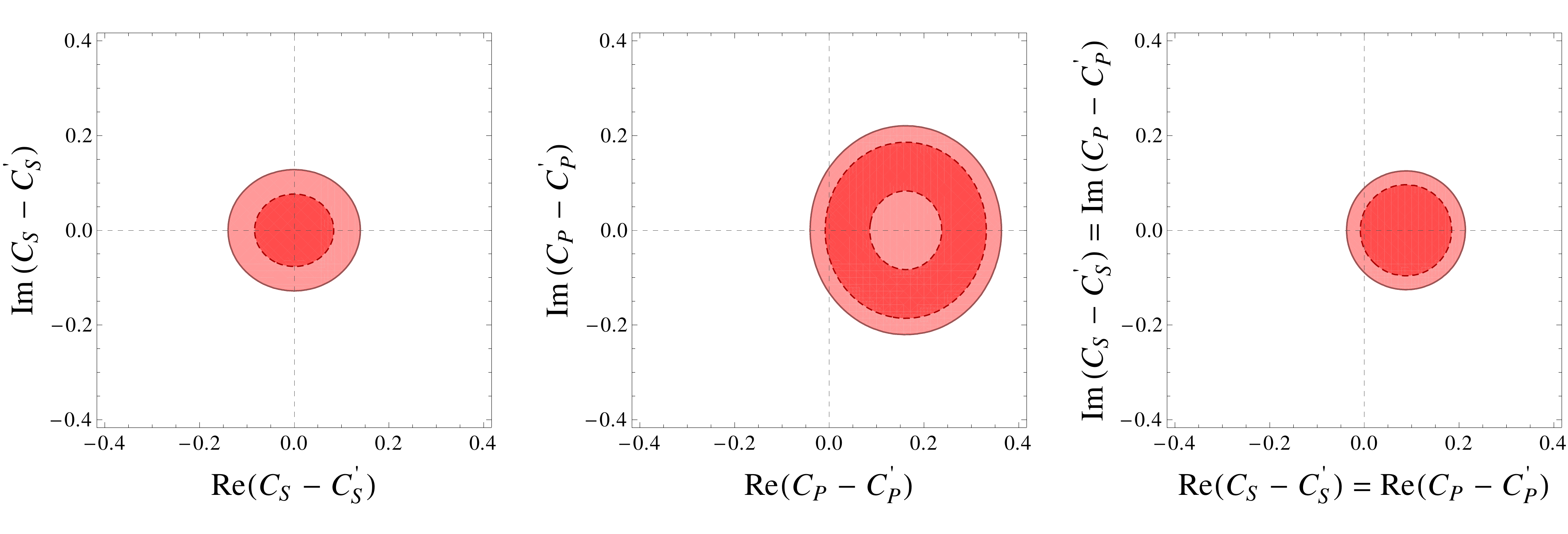}
\caption{Constraints on the scalar Wilson coefficients $C_S^{(\prime)}$ and $C_P^{(\prime)}$ at the 1 and $2\sigma$ level from BR$(B_s \to \mu^+ \mu^-)$, assuming no new physics in $C_{10}^{(\prime)}$.
}
\label{fig:bandplots_scalar}
\end{figure}

The new experimental bounds on BR($\bsmm$) can be used to constrain the combinations of scalar and pseudoscalar Wilson coefficients $C_{S,P}-C_{S,P}'$. The resulting bounds are shown in figure~\ref{fig:bandplots_scalar}, assuming $C_{10}^{(\prime)}$ to be SM-like. The right plot showing the constraint for $C_S - C_S^\prime = C_P - C_P^\prime$ can be easily reinterpreted in the frequently studied framework of the minimal flavor violating MSSM in the large $\tan\beta$ regime where $C_S \simeq - C_P$ and $C_S^\prime \simeq C_P^\prime \simeq C_{10} \simeq C_{10}^\prime \simeq 0$.
Complementary bounds on the combinations $C_{S,P}+C_{S,P}'$ can be obtained from the $B\to K\mu^+\mu^-$ decay, as has been done recently \cite{Becirevic:2012fy}. However, given the strong bound from $\bsmm$, this is only relevant if one considers $C_{S,P}+C_{S,P}'\gg C_{S,P}-C_{S,P}'$.

\begin{table}[tbp]
\renewcommand{\arraystretch}{1.4}
\small
\begin{center}
\begin{tabular}{rlcccccccc}
\hline
\multicolumn{2}{c}{Operator}& \multicolumn{4}{c}{$\Lambda$ [TeV] for $|c_i|=1$}
& \multicolumn{4}{c}{$|c_i|$ for $\Lambda=1$ TeV}\\
&& $+$ & $-$ & $+i$ & $-i$ &  $+$ & $-$ & $+i$ & $-i$\\
\hline
$\mathcal O_7 = $&$\frac{m_b}{e}(\bar{s} \sigma_{\mu \nu} P_R b) F^{\mu \nu}$&
 69 & 270 & 43 & 38 & $1.6 \cdot 10^{-4}$ & $9.7 \cdot 10^{-6}$ & $4.2 \cdot 10^{-4}$ & $5.3 \cdot 10^{-4}$ \\
$\mathcal O_7'= $&$\frac{m_b}{e}(\bar{s} \sigma_{\mu \nu} P_L b) F^{\mu \nu}$ & 
 46 & 70 & 78 & 47 & $3.6 \cdot 10^{-4}$ & $1.6 \cdot 10^{-4}$ & $1.3 \cdot 10^{-4}$ & $3.5 \cdot 10^{-4}$ \\
 $\mathcal O_9 = $&$(\bar{s} \gamma_{\mu} P_L b)(\bar{\ell} \gamma^\mu \ell)$ &
 29 & 64 & 21 & 22 & $1.2 \cdot 10^{-3}$ & $2.4 \cdot 10^{-4}$ & $2.2 \cdot 10^{-3}$ & $2.0 \cdot 10^{-3}$ \\
 $\mathcal O_9'= $&$(\bar{s} \gamma_{\mu} P_R b)(\bar{\ell} \gamma^\mu \ell)$ &
 51 & 22 & 21 & 23 & $3.8 \cdot 10^{-4}$ & $2.1 \cdot 10^{-3}$ & $2.2 \cdot 10^{-3}$ & $1.9 \cdot 10^{-3}$ \\
 $\mathcal O_{10} = $&$(\bar{s} \gamma_{\mu} P_L b)( \bar{\ell} \gamma^\mu \gamma_5 \ell)$ &
 43 & 33 & 23 & 23 & $5.4 \cdot 10^{-4}$ & $9.2 \cdot 10^{-4}$ & $1.9 \cdot 10^{-3}$ & $1.9 \cdot 10^{-3}$ \\
 $\mathcal O_{10}' = $&$(\bar{s} \gamma_{\mu} P_R b)( \bar{\ell} \gamma^\mu \gamma_5 \ell)$ &
 25 & 89 & 24 & 23 & $1.7 \cdot 10^{-3}$ & $1.3 \cdot 10^{-4}$ & $1.7 \cdot 10^{-3}$ & $1.9 \cdot 10^{-3}$ \\
 $\mathcal O_S^{(\prime)} = $&$\frac{m_b}{m_{B_s}} (\bar{s} P_{R(L)} b)(  \bar{\ell} \ell)$ &
 93 & 93 & 98 & 98 & $1.1 \cdot 10^{-4}$ & $1.1 \cdot 10^{-4}$ & $1.1 \cdot 10^{-4}$ & $1.1 \cdot 10^{-4}$ \\
 $\mathcal O_P = $&$\frac{m_b}{m_{B_s}} (\bar{s} P_R b)(  \bar{\ell} \gamma_5 \ell)$ &
 173 & 58 & 93 & 93 & $3.3 \cdot 10^{-5}$ & $3.0 \cdot 10^{-4}$ & $1.1 \cdot 10^{-4}$ & $1.1 \cdot 10^{-4}$ \\
 $\mathcal O_P' = $&$\frac{m_b}{m_{B_s}} (\bar{s} P_L b)(  \bar{\ell} \gamma_5 \ell)$ &
 58 & 173 & 93 & 93 & $3.0 \cdot 10^{-4}$ & $3.3 \cdot 10^{-5}$ & $1.1 \cdot 10^{-4}$ & $1.1 \cdot 10^{-4}$ 
\\\hline
\end{tabular}
\end{center}
\caption{Lower bounds (at 95\% C.L.) on the NP scale related to the relevant dimension six operators, assuming the coefficient $c_i$ in the effective Hamiltonian $\mathcal H_\text{eff}=-c_i\mathcal O_i/\Lambda^2$ to be $+1$, $-1$, $+i$ or $-i$ as well as upper bounds on the coefficients $c_i$ for a NP scale of $\Lambda = 1$~TeV, depending on $c_i/|c_i|$.}
\label{tab:npscale} 
\end{table}

The constraints on the individual Wilson coefficients can be translated into bounds on the suppressing scale of the flavour-violating dimension-six operators.
Table~\ref{tab:npscale} shows the constraints on the scales $\Lambda_i$ in the effective Hamiltonian\footnote{The constraints on the $c_i$ can be easily translated to constraints on the $C_i$ defined in eq.~\ref{eq:Heff} by dividing them by $(1\,\text{TeV})^2\frac{4\,G_F}{\sqrt{2}} V_{tb}V_{ts}^* \frac{e^2}{16\pi^2}\approx8.2\times10^{-4}$.} %
\begin{equation}
\mathcal H_\text{eff}=-\sum_{i={7,9,10,S,P}}\left(\frac{c_i}{\Lambda^2}\,\mathcal O_i+\frac{c_i'}{\Lambda^2}\,\mathcal O_i'\right)+\text{ h.c.}\,,
\end{equation}
assuming $|c_i|=1$, as well as the constraints on the $c_i$, assuming  $\Lambda=1$~TeV.
The constraints were obtained by varying only one coefficient at a time and correspond to $\Delta \chi^2=4$.
In both cases, we show the constraints for $c_i/|c_i|=+1,-1,+i$ or $-i$.

On average, the obtained constraints are slightly weaker compared to the constraints that one finds for the suppressing scales of dimension-six $\Delta F = 2$ operators that lead to $B_s$ mixing~\cite{Isidori:2010kg}.
Exeptions are the magnetic operator if it interferes with the SM destructively in $B \to X_s \gamma$ or scalar operators if they interefere constructively in $B_s \to \mu^+\mu^-$. Those operators are probed to scales above 100 TeV, as are operators that induce $B_s$ mixing.

\section{Allowed effects and future prospects}\label{sec:prospects}

The constraints in the previous section were derived assuming only one Wilson coefficient at a time, or the real parts of two coefficients, to deviate simultaneously from the SM. In the generic case, where all Wilson coefficients are allowed to deviate from the SM, cancellations may occur which render some of the constraints ineffective. On the other hand, even if one takes into account such cancellations, the current data already put indirect limits on observables which have not been measured to a good precision yet.

To obtain such predictions for allowed regions in the presence of generic NP, we perform a Bayesian Markov Chain Monte Carlo (MCMC) analysis of
six different scenarios where subsets of the Wilson coefficients or all of them have been varied, assuming flat priors for the real and imaginary parts of the Wilson coefficients, and using $e^{-\chi^2/2}$ as likelihood, with the $\chi^2$ function described above\footnote{In obtaining the predictions in table~\ref{tab:pred}, we have not included $A_\text{CP}(b\to s\gamma)$ in the $\chi^2$ function, since $C_7^{(\prime)}$ and $C_8^{(\prime)}$ enter this observable in a different linear combination compared to all other observables, so in a generic fit of Wilson coefficients, the constraint could always be compensated by adjusting $C_{7,8}^{(\prime)}$.} (for more details see \cite{Altmannshofer:2011gn}).
We then determine the Bayesian posterior credibility regions for each observable (for central values of the theory parameters) in each scenario.
These ranges can be interpreted as allowed regions in the different scenarios (with or without right-handed currents, with or without CPV beyond the CKM phase) and give an indication of the prospects of future measurements of these observables.

In table~\ref{tab:pred}, these fit predictions are shown at 95\% C.L. for the branching ratio of $\bsmm$ and for various angular observables in $B\to K^*\mu^+\mu^-$ sensitive to NP. 
We observe in particular that
\begin{itemize}
\item a suppression of BR($\bsmm$) below $10^{-9}$ requires NP in both left- and right-handed currents, i.e. in both $C_{10}$ and $C_{10}'$.\footnote{We remind the reader that here we assume (pseudo-)scalar currents to be absent.}
\item The T-odd CP asymmetries $A_{7,8,9}$ can still exceed 10\% (for $A_7$ even 30\%) in the presence of non-standard CPV. For effects in $A_8$ at high $q^2$ and $A_9$, right-handed currents are required.
\end{itemize}

While $A_8$ and $A_9$ at high $q^2$ are null tests of the SM and therefore promising inspite of sizable relative uncertainties, we point out that $S_3$ at high $q^2$ is non-zero even in the SM and afflicted with a large uncertainty (cf. table~\ref{tab:obs}), so that the large variation of the central value in the presence of right-handed currents shown in the last line of table~\ref{tab:pred} is spoiled by the badly known SM value. This would change if the relevant form factors could be estimated more precisely, e.g. by means of lattice QCD.

\begin{table}[tp] 
\renewcommand{\arraystretch}{1.6}
\begin{center}
\small
\begin{tabular}{ccccccc}
\hline
$C_i$                          & $\mathbbm{R}$ & $\mathbbm{C}$ &               &               & $\mathbbm{R}$ & $\mathbbm{C}$ \\
$C_i'$                         &               &               & $\mathbbm{R}$ & $\mathbbm{C}$ & $\mathbbm{R}$ & $\mathbbm{C}$ \\
\hline
$10^{9}\,\text{BR}(B_s\to\mu\mu)$ & $[1.9,5.2]$   & $[1.1,4.6]$   & $[1.1,4.2]$   & $[0.9,4.6]$   & $<4.6$   & $<4.2$   \\
$|\langle A_7\rangle_{[1,6]}|$ & $0$        & $<0.28$        & $0$        & $<0.22$        & $0$        & $<0.35$        \\
$|\langle A_8\rangle_{[1,6]}|$ & $0$        & $<0.14$        & $0$        & $<0.15$        & $0$        & $<0.21$        \\
$|\langle A_8\rangle_{[14.18,16]}|$ & $0$        & $0$        & $0$        & $<0.13$        & $0$        & $<0.12$        \\
$|\langle A_9\rangle_{[1,6]}|$ & $0$        & $0$        & $0$        & $<0.09$        & $0$        & $<0.13$        \\
$|\langle A_9\rangle_{[14.18,16]}|$ & $0$        & $0$        & $0$        & $<0.22$        & $0$        & $<0.20$        \\
$\langle S_3\rangle_{[1,6]}$ & $0$        & $0$        & $[-0.02,0.08]$        & $[-0.04,0.09]$        & $[-0.06,0.15]$        & $[-0.07,0.13]$        \\
$\langle S_3\rangle_{[14.18,16]}$ & $-0.14$        & $-0.14$        & $[-0.16,0.01]$        & $[-0.17,0.03]$        & $[-0.18,0.01]$        & $[-0.17,0]$        \\
\hline
\end{tabular}
\end{center}
\caption{Predictions at 95\% C.L. for the branching ratio of $\bsmm$  and predictions
for angular observables in $B\to K^*\mu^+\mu^-$ (neglecting tiny SM effects
below the percent level).
The columns correspond to 6 scenarios with real ($\mathbbm R$) or complex ($\mathbbm C$) new physics contributions to the operators $C_{7,9,10}$ and/or $C_{7,9,10}'$.
}
\label{tab:pred} 
\end{table}

\begin{figure}[tbp]
\centering
\includegraphics[width=0.35\textwidth]{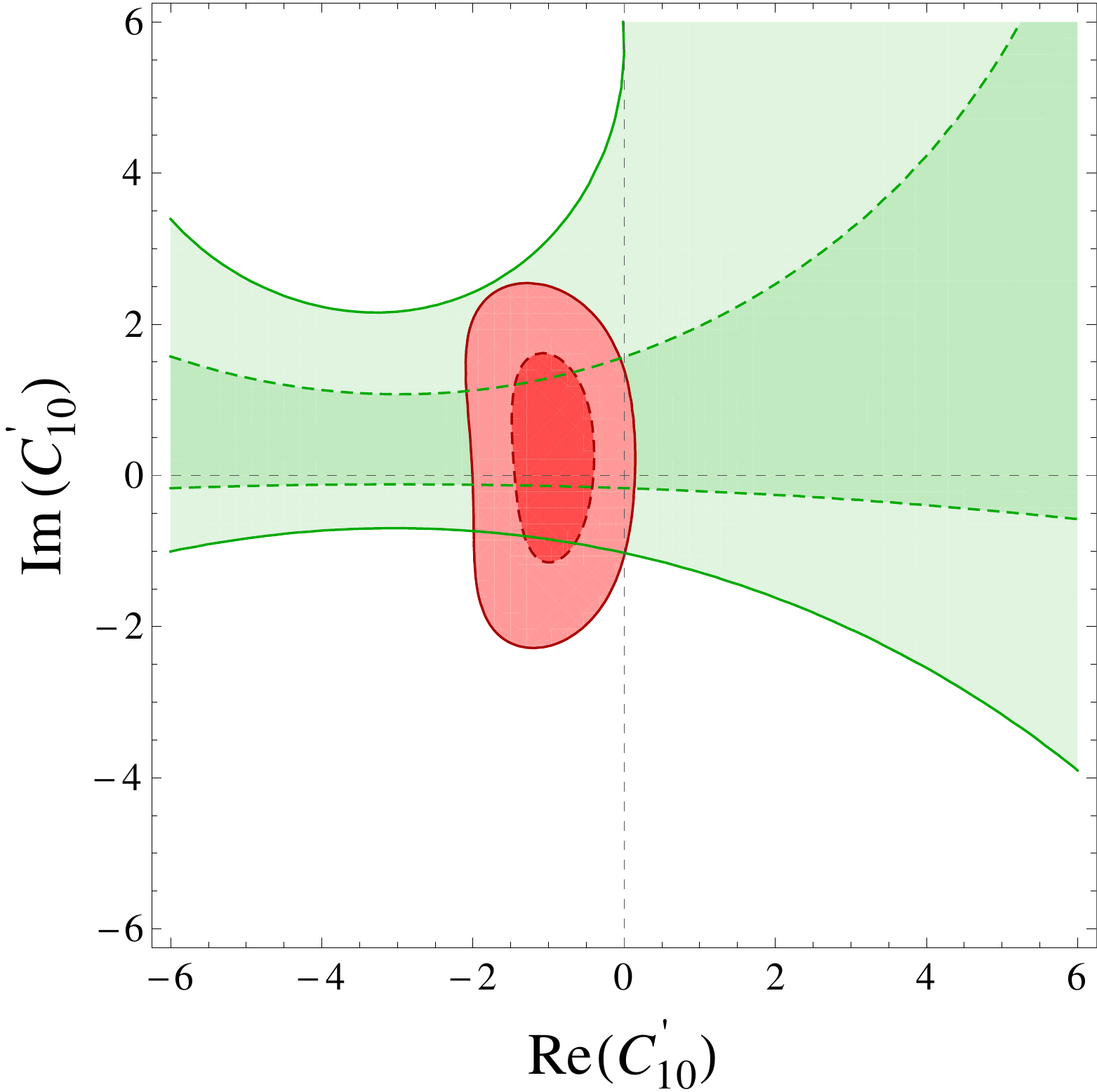}
\caption{Constraint on $C_{10}'$ at 68\% (green dashed) and 95 \% C.L. (green solid) from a hypothetical measurement of $A_9$ with the SM central value and the experimental errors equal to the current LHCb errors on $S_9$, compared to the current global constraint on $C_{10}'$ (red).}
\label{fig:bandplot_A9}
\end{figure}

Finally, we want to emphasize the importance of measuring the angular CP asymmetry $A_9$ in $B\to K^*\mu^+\mu^-$. As is well known, $A_9$ is highly sensitive to NP in right-handed currents at low $q^2$, and we stressed above that this is also true for the high $q^2$ region (cf.\ table~\ref{tab:sens}). Moreover, $A_9$ can be obtained from a merely one-dimensional angular distribution. Using the conventions of \cite{Altmannshofer:2008dz}, the distribution in the angle $\phi$ reads
\begin{equation}
\frac{d(\Gamma+\bar\Gamma)}{d\phi \,dq^2} \bigg/ \frac{d(\Gamma+\bar\Gamma)}{dq^2}
=
\frac{1}{2\pi} \left[
 1  +
 S_3  \cos(2\phi) +
 A_9  \sin(2\phi)
\right]
\,,
\label{eq:1Dphi}
\end{equation}
that is, $A_9$ can be extracted from an untagged sample \cite{Bobeth:2008ij} just as $S_3$.
Indeed, in 2011 CDF presented the first measurement of $A_9$ \cite{Aaltonen:2011ja}, denoted $A_\text{Im}$ by CDF. The preliminary results for the quantity also denoted $A_\text{Im}$ recently presented by LHCb \cite{LHCb-CONF-2012-008} correspond instead\footnote{Thomas Blake and Nicola Serra, private communication.} to the CP-averaged angular observable $S_9$, whose sensitivity to NP is very limited \cite{Altmannshofer:2008dz}.  To demonstrate the constraining power of $A_9$ on CP-violating right-handed currents, we show in figure~\ref{fig:bandplot_A9} the constraints from a hypothetical measurement of $A_9$ with SM central value (i.e. zero), and the experimental errors equal to the current errors of LHCb's measurement of $S_9$. As the figure shows, such measurement, which should be possible even with the current LHCb dataset, would significantly reduce the allowed size of NP in $\text{Im}(C_{10}')$.

\section{Conclusions}\label{sec:concl}

Using the recent improved measurements of rare decays probing the $b\to s$ transition, we have derived updated constraints on Wilson coefficients of dimension-six effective operators. We went beyond comparable recent studies by including the most complete set of experimental observables and NP operators, including also the chirality-flipped counterparts of the operators present in the SM. Compared to the predecessor study  \cite{Altmannshofer:2011gn}, in addition to including the new experimental data, we made several improvements, the most significant of which being the inclusion of the decays $B\to K\mu^+\mu^-$ and $\bsmm$ as well the direct CP asymmetry in $b\to s\gamma$ in our constraints.

The most important results can be summarized as follows.
\begin{itemize}
\item At the 95\% C.L., the Wilson coefficients are still compatible with their SM values.
\item The branching ratio of $B\to K\mu^+\mu^-$ constitutes a particularly important constraint on the chirality-flipped Wilson coefficients $C_{9,10}'$, for which it is complementary to $B\to K^*\mu^+\mu^-$.
\item The new LHCb measurement of the angular observable $S_3$ in $B\to K^*\mu^+\mu^-$ constitutes a significant constraint on right-handed currents. It is sensitive to NP also in the high $q^2$ region.
\item A measurement of the CP asymmetry $A_9$ is expected to provide a significant constraint on Im$(C_{9,10}')$ even with the current LHCb dataset, if measured to be compatible with the SM.
\item In the high-$q^2$ region of $B\to K^*\mu^+\mu^-$, there are five angular observables only sensitive to right-handed currents: $F_L$, $S_3$, $S_4$, $A_8$ and $A_9$.
\end{itemize}

In the near future, improved constraints -- or chances to uncover physics beyond the SM! -- will be facilitated by improved measurements of the branching ratios of $B\to K\mu^+\mu^-$, $B\to X_s\ell^+\ell^-$ and $B_s \to \mu^+\mu^-$ and the remaining CP-symmetric and asymmetric angular observables in $B\to K^*\mu^+\mu^-$. As we have also seen, several observables in exclusive decays are already dominated by theory uncertainties, so progress in lattice calculations of hadronic form factors would be very welcome.

\section*{Acknowledgements}

We acknowledge Aoife Bharucha, Thomas Blake, Nicola Serra and Liang Sun for useful discussions.
Fermilab is operated by Fermi Research Alliance, LLC under Contract No. De-AC02-07CH11359 with
the United States Department of Energy. D.M.S. is supported by the EU ITN ``Unification in the
LHC Era'', contract PITN-GA-2009-237920 (UNILHC).

\appendix
\section{Averaging procedure and theory vs. experiment in $B\to K^{(*)}\mu^+\mu^-$}\label{app:exp}

\begin{figure}[tbp]
\centering
\includegraphics[width=\textwidth]{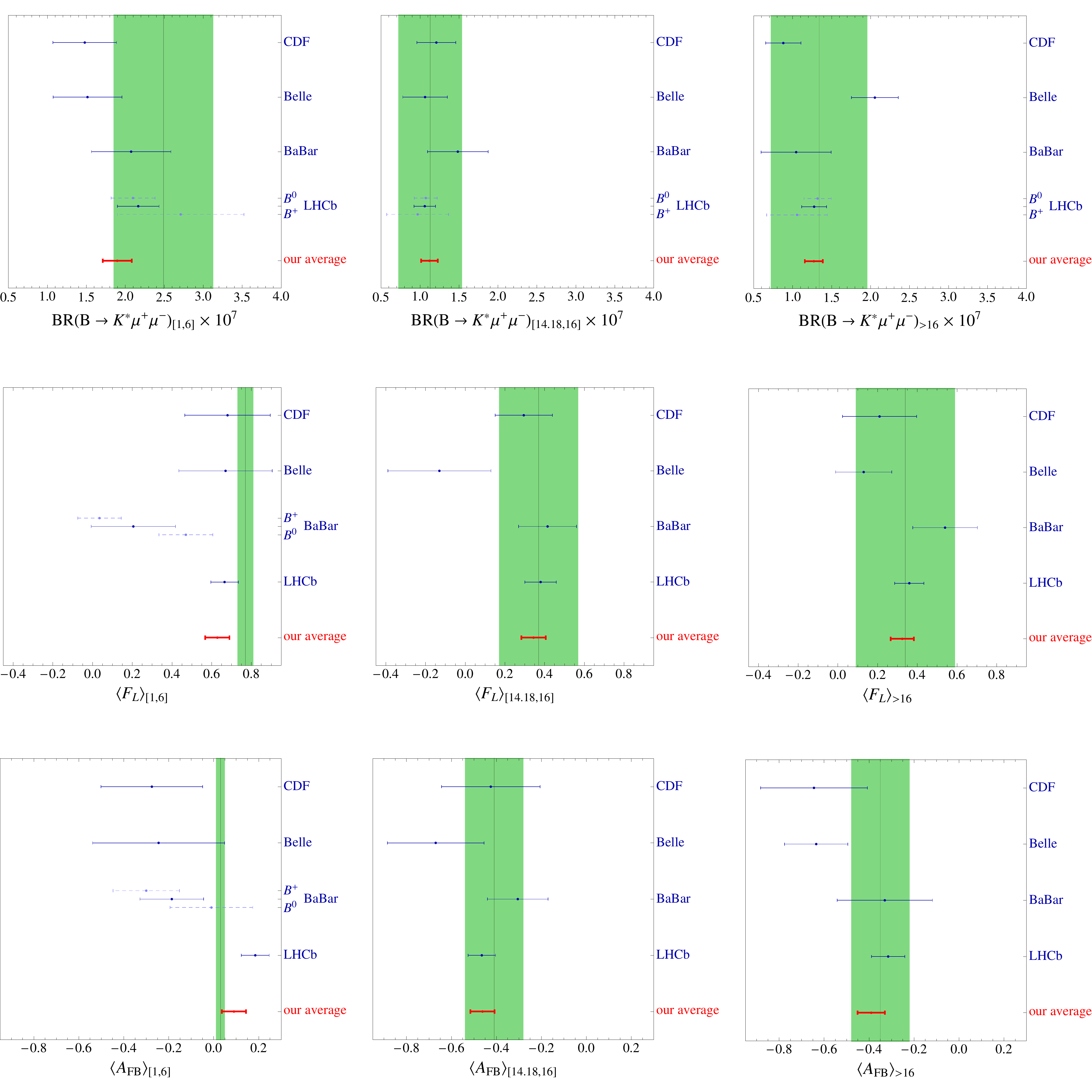}
\caption{Symmetrized experimental results and averages obtained using the procedure described in appendix~\ref{app:exp} for the $B\to K^*\mu^+\mu^-$ branching ratio, $F_L$ and $A_\text{FB}$ in the three $q^2$ regions considered. The green bands are our theory predictions with $1\sigma$ uncertainties. In the cases where we perform the average of $B^0$ and $B^+$ data ourselves (for details on the procedure see text), we also show the separate results as dashed error bars.}
\label{fig:Combination_Kstar}
\end{figure}

\begin{figure}[tbp]
\centering
\includegraphics[width=\textwidth]{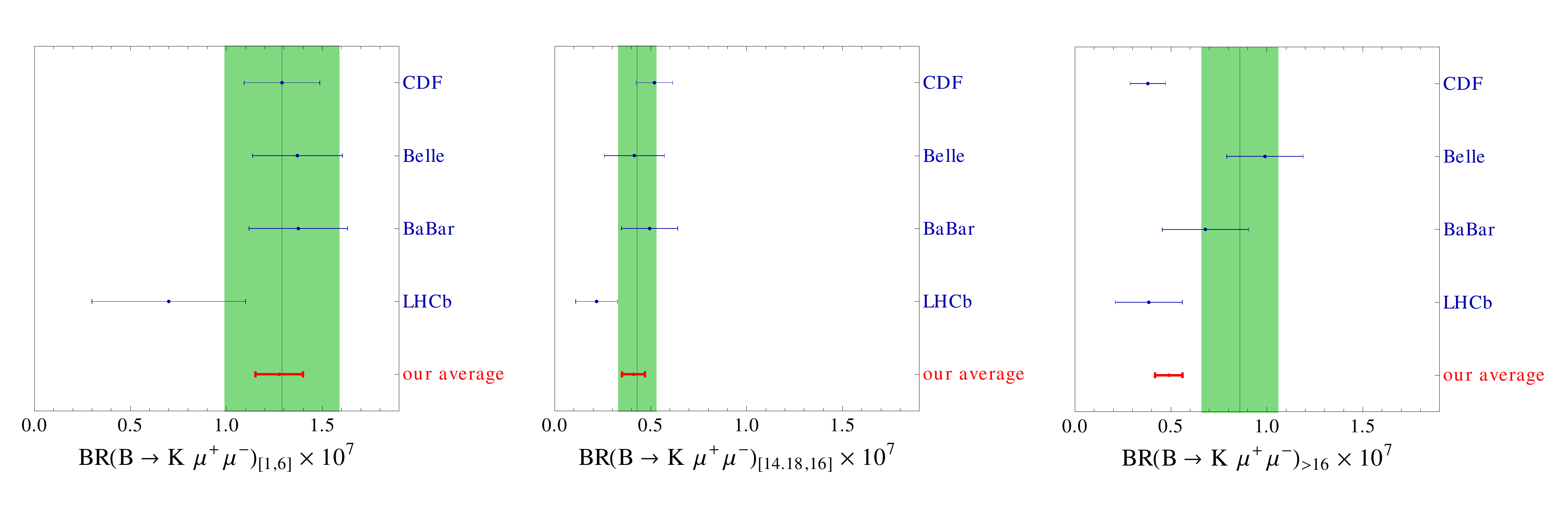}
\caption{Symmetrized experimental results and averages obtained using the procedure described in appendix~\ref{app:exp} for the $B\to K\mu^+\mu^-$ branching ratio in the three $q^2$ regions considered.  The green bands are our theory predictions with $1\sigma$ uncertainties.}
\label{fig:Combination_K}
\end{figure}

In this appendix, we describe our method to obtain the averages of experimental measurements of $B\to K^*\mu^+\mu^-$ and $B\to K\mu^+\mu^-$ observables quoted in table~\ref{tab:obs}.
Given a number of measurements of the same experimental quantity (like a partial branching fraction), we first symmetrize asymmetric statistical and/or systematic errors using the prescription of ref.~\cite{D'Agostini:2004yu}. Then, we perform a weighted average of the symmetrized individual results.
We use a slightly modified procedure in the case where experiments quote separate results for the charged and neutral $B\to K^{(*)}\mu^+\mu^-$ decays. As mentioned in section~\ref{sec:BKstar}, we assume the isospin asymmetry to be small with respect to the experimental uncertainties in all observables. However, in some cases, in particular the BaBar $B\to K^*\mu^+\mu^-$ data at low $q^2$, the results for the charged and neutral modes show a significant difference. In these cases, we first perform the average of the charged and neutral mode results of a single experiment using the PDG averaging method, i.e. rescaling the error by a factor of $\sqrt{\chi^2}$, and use this rescaled error in the weighted average with the other experiments.
If the charged and neutral mode results agree within experimental uncertainties or if the experimental sensitivity to one mode is much better than to the other, we use the combination provided in the experimental publications, if available. Our SM predictions for the $B\to K^{(*)}\mu^+\mu^-$ branching ratios in table~\ref{tab:obs} have been obtained using the neutral $B$ lifetime, so in the combination of experimental results, we have rescaled the charged $B$ branching ratios, if necessary.

Figure~\ref{fig:Combination_Kstar} confronts the individual (appropriately symmetrized) experimental measurements of the branching ratio, $F_L$ and $A_\text{FB}$ in $B\to K^*\mu^+\mu^-$ and our average to the SM prediction. We highlight the following points,
\begin{itemize}
\item The theory uncertainties are particularly large for the branching ratio and the high-$q^2$ angular observables. In these cases, the uncertainty is already dominated by theory and progress in constraining NP will only be possible with better control over the form factors, which might come from lattice QCD.
\item In the low-$q^2$ angular observables, there is some tension between BaBar and LHCb results, in addition to the tension among charged and neutral $B$ results observed by both experiments.
\end{itemize}

Figure~\ref{fig:Combination_K} shows the same comparison for the $B\to K\mu^+\mu^-$ branching ratio. Also here, the uncertainties are dominated by theory and require progress on the lattice calculation of form factors. We also observe a tension in the highest $q^2$ bin at a level of somewhat less than $2\sigma$, which however does not have a big impact on the constraints since the adjacent bin agrees well with the SM expectation and NP typically affects both bins simultaneously.

\bibliographystyle{utphys}
\bibliography{bsll}

\end{document}